\documentclass[a4paper, UKenglish, cleveref, autoref, thm-restate, unicode, colorlinks=true, allcolors=blue]{lipics-v2021}

\nolinenumbers{}

\usepackage{amsmath,amssymb}
\usepackage{graphicx}
\usepackage{xspace}
\usepackage{xcolor}
\usepackage{thmtools}
\usepackage{url}
\usepackage{svg}
\usepackage[T1]{fontenc}

\newtheorem{problem}{Problem}
\newtheorem{fact}{Fact}

\newcommand{\return}{\vspace{0.125cm}}

\def\dd{\mathinner{.\,.}} 
\newcommand{\cO}{\mathcal{O}}

\newcommand{\Minimizers}{\mathcal{M}_{\ell,k}\xspace}
\newcommand{\Size}{z}
\newcommand{\ctO}{\mathcal{\tilde{O}}}

\newcommand{\method}[1]{\textup{\textsf{#1}}}
\newcommand{\INDEX}{\method{Index}\xspace}
\newcommand{\TRIE}{\method{Trie}}
\newcommand{\Sketch}{\method{Sketch}}
\newcommand{\Locate}{\method{Locate}}
\newcommand{\Count}{\method{Count}}
\newcommand{\Extract}{\method{Extract}}
\newcommand{\Map}{\method{Map}}
\newcommand{\uindex}{\method{U-index}\xspace}
\newcommand{\sindex}{\method{sparse SA}\xspace}
\newcommand{\sa}{\method{SA}\xspace}
\newcommand{\fmindex}{\method{FM-index}\xspace}

\newcommand{\kmer}{$k$-mer\xspace}

\title{U-index: A Universal Indexing Framework for Matching Long Patterns}

\author{Lorraine A. K. Ayad}
       {Brunel University London, London, UK}
       {lorraine.ayad@brunel.ac.uk}
       {orcid.org/0000-0003-0846-2616}
       {}
\author{Gabriele Fici}
       {Dipartimento di Matematica e Informatica, Università di Palermo, Italy}
       {gabriele.fici@unipa.it}
       {https://orcid.org/0000-0002-3536-327X}
       {Supported by MUR project PRIN 2022 APML – 20229BCXNW, funded by the European Union – Mission 4 ``Education and Research'' C2 - Investment 1.1. CUP Master$\_$B53D23012910006.}
\author{Ragnar Groot Koerkamp}
       {ETH Zurich, Zurich, Switzerland}
       {ragnar.grootkoerkamp@inf.ethz.ch}
       {https://orcid.org/0000-0002-2091-1237}
       {ETH Research Grant ETH-1721-1 to Gunnar Rätsch.}
\author{Grigorios Loukides}
       {King's College London, London, UK}
       {grigorios.loukides@kcl.ac.uk}
       {https://orcid.org/0000-0003-0888-5061}
       {}
\author{Rob Patro}
       {University of Maryland, College Park, MD, USA}
       {rob@cs.umd.edu}
       {https://orcid.org/0000-0001-8463-1675}
       {NIH grant award number R01HG009937, NSF award CNS-1763680 and grants
         252586 and 2024342821 from the Chan Zuckerberg Initiative DAF, an
         advised fund of Silicon Valley Community Foundation.
         RP is a co-founder of Ocean Genomics, Inc.
       }
\author{Giulio Ermanno Pibiri}
       {Ca' Foscari University of Venice, Venice, Italy\\ISTI-CNR, Pisa, Italy}
       {giulioermanno.pibiri@unive.it}
       {https://orcid.org/0000-0003-0724-7092}
       {European Union’s Horizon Europe research and innovation programme (EFRA project, Grant Agreement Number 101093026). This work was also partially supported by DAIS – Ca’ Foscari University of Venice within the IRIDE program.}
\author{Solon P. Pissis}
       {CWI, Amsterdam, The Netherlands\\Vrije Universiteit, Amsterdam, The Netherlands}
       {solon.pissis@cwi.nl}
       {https://orcid.org/0000-0002-1445-1932}
       {Supported by the PANGAIA and ALPACA projects that have received funding from the European Union’s Horizon 2020 research and innovation programme under the Marie Skłodowska-Curie grant agreements No 872539 and 956229, respectively.}

\authorrunning{L.A.K.\ Ayad, et. al.} 
\Copyright{Lorraine A. K. Ayad, Gabriele Fici,  Ragnar Groot Koerkamp, Grigorios Loukides, Rob Patro, Giulio Ermanno Pibiri, Solon P. Pissis}

\ccsdesc{Theory of computation~Sketching and sampling}
\ccsdesc{Applied computing~Bioinformatics}

\supplement{}
\supplementdetails[subcategory={Rust}, cite={}, swhid={}]{Software}{https://github.com/u-index/u-index-rs}

\relatedversiondetails[cite={u-index-arxiv-version}]{Full version}{https://arxiv.org/abs/2502.14488}

\keywords{Text Indexing; Sketching; Minimizers; Hashing}

\EventEditors{Petra Mutzel and Nicola Prezza}
\EventNoEds{2}
\EventLongTitle{23rd International Symposium on Experimental Algorithms (SEA 2025)}
\EventShortTitle{SEA 2025}
\EventAcronym{SEA}
\EventYear{2025}
\EventDate{July 22--24, 2025}
\EventLocation{Venice, Italy}
\EventLogo{}
\SeriesVolume{338}
\ArticleNo{27}

\begin{document}
\maketitle

\begin{abstract}

\subparagraph{Motivation.}
Text indexing is a fundamental and well-studied problem.
Classic solutions to this problem either replace the original text with a compressed representation, e.g., the FM-index and its variants, or keep it uncompressed but attach some redundancy --- an index --- to accelerate matching, e.g., the suffix array. The former solutions thus retain excellent compressed space, but are practically slow to construct and query. The latter approaches, instead, sacrifice space efficiency but are typically faster; for example, the suffix array takes much more space than the text itself for commonly used alphabets, like ASCII or DNA, but it is very fast to construct and query.

\subparagraph{Methods.}
In this paper, we show that efficient text indexing
can be achieved using just a small extra space on top of the original text, provided that the query patterns are sufficiently long.
More specifically, we develop a new indexing paradigm in which a \emph{sketch} of a query pattern is first matched against a \emph{sketch} of the text. Once \emph{candidate} matches are retrieved, they are verified using the original text.
This paradigm is thus \textit{universal} in the sense that it allows us to use
\textit{any} solution to index the sketched text, like a suffix array, FM-index, or r-index.

\subparagraph{Results.}
We explore both the theory and the practice of this universal framework. With an extensive experimental analysis, we show that, surprisingly, universal indexes can be constructed much faster than their unsketched counterparts and take a fraction of the space, as a direct consequence of (i) having a lower bound on the length of patterns and (ii) working in sketch space. 
Furthermore, these data structures have the potential of retaining or even improving query time, because matching against the sketched text is faster and verifying candidates can be theoretically done in constant time per occurrence (or, in practice, by short and cache-friendly scans of the text).

Finally, we discuss some important applications of this novel indexing paradigm to computational biology. We hypothesize that such indexes will be particularly effective when the queries are sufficiently long, and so we demonstrate applications in long-read mapping.

\end{abstract}

\setcounter{page}{1}

\section{Introduction}\label{sec:introduction}


The problem of \textit{text indexing}~\cite{navarro2007compressed}, at its core, involves finding all occurrences of a given pattern within a large body of text. Formally, the problem is as follows.

\begin{problem}[Text Indexing]
Given a string $T[0\dd n)$ (the ``text'' henceforth) of $n$ characters over an alphabet $\Sigma = [0 , \sigma)$, we are asked to pre-process $T$ so that the following queries are supported efficiently for any string $P[0 \dd m)$ (the ``pattern'' henceforth) and any $0 \leq i < j \leq n$:
\begin{itemize}
    \item $\Locate(P,T)$ determines the set $L = \{0 \leq i < n-m+1 \mid T[i \dd i+m) = P \}$;
    \item $\Count(P,T)$ returns $|L|$; and
    \item $\Extract(i,j,T)$ returns $T[i \dd j)$.
\end{itemize}
\end{problem}

Given the importance of text indexing, numerous classic solutions have been proposed, each with varying trade-offs between time and space efficiency. In general terms, these solutions fall into two categories: the \emph{compressed} and \emph{uncompressed} approaches.

A common strategy is to replace the original text with a compressed representation (a so-called ``self-index''), utilizing data structures like the \textit{FM-index}~\cite{10.1145/1082036.1082039} or its modern variants such as the \textit{r-index}~\cite{10.1145/3375890}. These indexes are highly space-efficient,  achieving space bounds in terms of the $k$-th order empirical entropy of the text, while retaining the ability to support searches. However, this efficiency in space comes at the cost of increased complexity in both the construction and query phases~\cite{navarro2007compressed}. Compressed indexes typically require more time to build, and querying can be slower compared to uncompressed counterparts due to the additional overhead involved in decoding the compressed representation.

On the other hand, uncompressed approaches, such as
the \textit{suffix array}~\cite{DBLP:journals/siamcomp/ManberM93}, do not alter the original text but instead attach some form of redundancy --- henceforth referred to as the \textit{index} --- to accelerate pattern matching. These methods, while being faster in terms of query times, often suffer from significant space inefficiency. For example, suffix arrays usually take up much more space than the text itself, particularly for commonly used alphabets such as ASCII or DNA~\cite{gusfield1997algorithms}. This makes such structures impractical for many applications, especially when working with very large datasets.

\subparagraph{Our Contributions.}
In this paper, we focus on the latter approach of uncompressed text indexing and
aim to address its space inefficiency while retaining efficient queries. In the
following, we discuss the {\Locate} query only, given that {\Count} can be
trivially implemented using {\Locate}, and {\Extract} is done by explicitly
accessing the text $T$. We propose a novel approach that solves the text
indexing problem using only a small amount of additional space beyond the
original text, without sacrificing the speed of query processing. This is
achieved under the assumption that the query patterns $P[0 \dd m)$ are
  sufficiently long, i.e., we require that $m \geq \ell$ for some fixed lower
  bound $\ell > 0$~\cite{DBLP:journals/pvldb/AyadLP23}, and operate under the assumption that $n \gg \ell$. 
A typical value for $\ell$ lies in $[32,1000]$. The core idea of our approach is to transform the text into sketch space, and to
construct and use a smaller index over the sketched text to enable queries with only small additional space.
Generally speaking, a sketch is a compact representation of an object that retains enough information for approximate matches. 

We therefore introduce a four-step framework for addressing text indexing:
\begin{enumerate}
  \item We sketch the text $T$, say $S=\Sketch(T)$. The sketch $S$ can be simply regarded as a new shorter string over a new alphabet $\Sigma'$.
  \item We then construct $\INDEX(S)$, an index on $S$, where $\INDEX$ is any indexing structure.
  \item Each query pattern $P$ is then sketched into $Q=\Sketch(P)$, and $Q$ is
    matched against $\INDEX(S)$ to identify \emph{potential} matches $L' =
    \Locate(Q, S)$.
  \item Candidate matches in $L'$ are mapped back to their positions in
    $T$, where we verify that indeed $P$
    matches $T$. Thus we obtain $L=\Map(L')$, where $L=\Locate(P,T)$.
\end{enumerate}

The \textit{universality} of our framework lies in its flexibility: (i) \textit{any} indexing structure, $\INDEX$, such as the FM-index or the suffix array, can be used to index the sketched text $S$, making the method adaptable to a variety of text indexing techniques~\cite{DBLP:journals/siamcomp/ManberM93,10.1145/1082036.1082039,grossi2005compressed,10.1145/3375890,DBLP:conf/soda/KempaK23,DBLP:conf/focs/KempaK23}; and (ii) \textit{any} locally consistent sampling mechanism, $\Sketch$, such as minimizers~\cite{DBLP:journals/bioinformatics/RobertsHHMY04,DBLP:conf/sigmod/SchleimerWA03}, syncmers~\cite{Syncmers}, or bd-anchors~\cite{DBLP:journals/tkde/LoukidesPS23},
can be used to sketch $S$ and $P$, making the method adaptable to a variety of sampling techniques.
We also expect that our framework, which relies on $\INDEX(S)$, has competitive queries compared to the ones of $\INDEX(T)$ or possibly faster when the query time of $\INDEX(T)$ is a function of $n=|T|$. This is merely because $|S|<|T|$ (in practice, depending on the sketching mechanism used, $|S|$ could be $\Theta(|T|/\ell)$ or $\Theta(|T|/(\ell-k))$ for some $k=o(\ell)$). A typical such case is when $\INDEX$ is the suffix array~\cite{DBLP:journals/siamcomp/ManberM93} and binary search will take place on a \emph{smaller array}, i.e., the suffix array of $S$ instead of the suffix array of $T$.

We explore both the theoretical underpinnings and the practical implementation
of this universal framework.
We demonstrate that universal indexes can be constructed significantly faster
and occupy a fraction of the space compared to their unsketched counterparts.
These space savings are a direct consequence of operating in the reduced
``sketch space''. Depending on the index used, the performance of pattern
matching in sketched space can be either faster (suffix array) or slower
(FM-index). In either case, the overall query performance is slightly slower
though, due to a relatively large number of false positive matches that have to
be rejected during verification.
This verification step can be performed in constant time
per occurrence (in theory) or via cache-friendly scans of the original
text (in practice).



\return
In short, the \emph{main message} of our paper is the following:
If we have a sufficiently large lower bound $\ell$ (e.g., 32 or more), then our universal scheme typically offers substantial improvements over any text index in construction time, construction space, and index size, while supporting competitive query times.

\subparagraph{Paper Organization.}
In \cref{sec:background}, we provide the necessary notation and tools.
In \cref{sec:related}, we provide an overview of previous related work.
In \cref{sec:framework}, we present our framework, and
in \cref{sec:experiments}, we present our experimental evaluation.
We conclude this paper in \cref{sec:conclusions}.

\section{Preliminaries}\label{sec:background}

In this section, we provide useful background information to support subsequent descriptions.

\subsection{Notation and Computational Model}

\subparagraph{Basic Notation.}
Recall that we deal with a string $T[0 \dd n)$ of length $n$ over an alphabet $\Sigma=[0,\sigma)$.
We assume that $\Sigma$ is an integer alphabet of polynomial size in $n$, i.e., $\sigma=n^{\cO(1)}$. 
A substring of $T$ of length $k>0$ is called a {\kmer}.
Given a query pattern $P[0 \dd m)$ with $m \geq \ell$ for some lower bound $\ell > 0$, the goal is to support the following operation:
$$\Locate(P,T) = \{0 \leq i < n-m+1 \, | \, T[i \dd i+m) = P \}.$$
We refer to the number of occurrences of $P$ in $T$, that is, $|\Locate(P,T)|$, with $\Count(P,T)$. 


\subparagraph{The Computational Model.}
We assume that we have random read-only access to $T$ and
count the space (in number of words) occupied on top of the space occupied by $T$. We assume the standard word RAM model of computation with machine words of $\Omega(\log n)$ bits.

\subsection{Algorithmic Toolbox}

The solutions we describe in this paper rely on few, well-defined tools that we present below.

\subparagraph{Minimizers.} We use a specific class of randomized methods to sketch a string, called \textit{minimizers}~\cite{DBLP:journals/bioinformatics/RobertsHHMY04,DBLP:conf/sigmod/SchleimerWA03}. Minimizers are defined as the triple $(k,w,h)$: from a window of $w$ consecutive {\kmer}s of $T$, the leftmost smallest {\kmer} according to the order $h$ is chosen as the \textit{minimizer} of the window.
Since at least one {\kmer} must be chosen every $w$ positions, the fraction of sampled {\kmer}s --- defined as the \textit{density} of the sampling algorithm --- is always at least $1/w$. Several minimizer sampling algorithms have been proposed in the literature. See Section 3 of \cite{grootkoerkamp_et_al:LIPIcs.WABI.2024.11} for a recent overview of different sampling strategies and orders that lead to different densities. 
Here, however, we use the folklore \textit{random minimizer} sampling, which is as defined above and uses a pseudo-random hash function for the order $h$. 
We have the following result.

\begin{theorem}[Theorem 3 from \cite{zheng2020improved}]\label{thm:random-mini}
When $T$ is a string of i.i.d.~random characters and $k > (3+\varepsilon)\log_{\sigma}(w+1)$ for any $\varepsilon > 0$, the density of the random minimizer is $2/(w+1) + o(1/w)$.
\end{theorem}

We fix $\ell$ to be the minimum pattern length and let $w=\ell-k+1$. Each substring of length $\ell$ of $T$ therefore contains one minimizer.
(In practice, we expect to have $|P|\gg\ell$ and that the sketch of $P$ is a sequence of several minimizers.)
Further, we let $\Minimizers(T)$ indicate the sorted list of positions in $T$ of the minimizers of $T$.
Let $\Size = |\Minimizers(T)|$ be the number of minimizers.
By \Cref{thm:random-mini}, we have that $\Size \approx 2(|T|-\ell+1)/(\ell-k+2)$ in expectation. 


\subparagraph{Tries.}
Given a set $\mathcal{X}$ of strings over the alphabet $\Sigma$, a \emph{trie} $\TRIE(\mathcal{X})$ is a rooted tree whose nodes represent the prefixes of the strings in $\mathcal{X}$.
The edges of $\TRIE(\mathcal{X})$ are labeled by letters from $\Sigma$; the prefix corresponding to node $u$ is denoted by $\textsf{str}(u)$ and is given by the concatenation of the letters labeling the path (sequence of edges) from the root of $\TRIE(\mathcal{X})$ to $u$. The node $u$ is called the \emph{locus} of $\textsf{str}(u)$. 
The parent-child relationship in $\TRIE(\mathcal{X})$ is defined as follows: the root node is the locus of the empty string $\varepsilon$; and the
parent $u$ of another node $v$ is the locus of $\textsf{str}(v)$ without the last letter. This letter is the edge label of $(u,v)$. The order on $\Sigma$ induces an order on the edges outgoing from any node of the trie. A node $u$ is \emph{branching} if it has at least two children and \emph{terminal} if $\textsf{str}(u) \in \mathcal{X}$. 

A \emph{compacted trie} is obtained from $\TRIE(\mathcal{X})$ by removing all nodes except the root, the branching nodes, and the terminal nodes.
A compacted trie is thus a trie where all unary paths are collapsed into a single edge, labeled by the string obtained by concatenating all the letters of the edges in the unary path.
The compacted trie takes $\cO(|\mathcal{X}|)$ space provided that the edge labels are implicitly represented as pointers to fragments of strings in $\mathcal{X}$. Given the lexicographic order on $\mathcal{X}$ along with the lengths of the longest common prefixes between any two consecutive elements (in this order) of $\mathcal{X}$, one can compute $\TRIE(\mathcal{X})$ in $\cO(|\mathcal{X}|)$ time~\cite{DBLP:conf/cpm/KasaiLAAP01}.


\subparagraph{Rolling Hashing.}
Let $p$ be a prime number and choose $r \in [0,p)$ uniformly at random. 
The rolling hash value of $T[i \dd j]$ --- which we term \textit{fingerprint} --- is defined as~\cite{DBLP:journals/ibmrd/KarpR87}: 
$$
\phi(T[i\dd j]) := \sum^{j}_{k=i}T[k]r^{j-k}\bmod p.
$$
The adjective ``rolling'' refers to the way the hash value is updated incrementally as a fixed-length window slides through the string $T$.
The function $\phi$ allows one to compute the fingerprint of a window just knowing the fingerprint of the previous window and the character that is being removed/added, instead of recalculating the fingerprint from scratch.

By definition, if $T[i\dd i + \ell] = T[j\dd j + \ell]$, then $\phi(T[i \dd i+\ell]) = \phi(T[j \dd j+\ell])$. On the other hand, if $T[i\dd i + \ell] \neq T[j\dd j + \ell]$, then $\phi(T[i \dd i+\ell]) \neq \phi(T[j \dd j+\ell])$ with probability at least $1-\ell/p$~\cite{DBLP:conf/icalp/DietzfelbingerGMP92}. Since we are comparing only substrings of equal length, the number of different possible substring comparisons is less than $n^3$. Thus, for any constant $c\geq 1$, we can set $p$ to be a
prime larger than $\max(|\Sigma|,n^{c+3})$ to make the function $\phi$ perfect (i.e., no collisions) with probability at least $1 - n^{-c}$ (this means \emph{with high probability}). Any fingerprint of $T$ or $P$ fits in one machine word, so that comparing any two fingerprints takes $\cO(1)$ time.
In particular, we will use the following well-known fact.

\begin{fact}[\cite{DBLP:journals/ibmrd/KarpR87}]\label{fact:KR}
For any $0 \leq i < j < n$, we have  
$$
\phi(T[i+1 \dd j]) = ( \phi(T[0 \dd j]) - r^{j-i}\phi(T[0 \dd i]) ) \bmod p.
$$
\end{fact}

\section{Related Work} \label{sec:related}

Text indexing for matching long patterns (i.e., with lengths at least $\ell$ for some $\ell > 0$) in the uncompressed setting
has attracted some attention in the literature~\cite{DBLP:journals/jda/ClaudeNPST12,DBLP:journals/spe/GrabowskiR17,DBLP:conf/esa/LoukidesP21,DBLP:journals/tkde/LoukidesPS23,DBLP:journals/pvldb/AyadLP23,DBLP:journals/corr/abs-2407-11819}. The common idea of these approaches is to use some form of sketching, such as alphabet sampling~\cite{DBLP:journals/jda/ClaudeNPST12}, minimizer-like anchors~\cite{DBLP:journals/spe/GrabowskiR17,DBLP:conf/esa/LoukidesP21,DBLP:journals/tkde/LoukidesPS23,DBLP:journals/pvldb/AyadLP23} or their worst-case counterparts~\cite{DBLP:journals/corr/abs-2407-11819}. The work of~\cite{DBLP:journals/jda/ClaudeNPST12} chooses a subset of the alphabet and constructs a sparse suffix array only for the suffixes starting with a letter from the chosen 
subalphabet. The search starts with finding the leftmost occurrence $j$ of a sampled letter of 
pattern $P$. Then the suffix $P[j\dd m)$ is sought
using the sparse suffix array with standard means. After that, each occurrence of
the suffix is verified against the text with the previous $j$ letters.
The work of~\cite{DBLP:journals/spe/GrabowskiR17} proposes a similar approach. It first computes the set $B$ of starting positions of the minimizers of text $T$
and then constructs the sparse suffix array only for the suffixes starting at the positions in $B$. Upon a query pattern $P$, it computes the starting position $j$ of the leftmost minimizer of $P$,
thus implicitly partitioning $P$ into $P[0\dd j)$ and $P[j\dd m)$. It then searches
$P[j\dd m)$ in the sparse suffix array, and verifies each occurrence of it using letter comparisons against $T$ using the previous $j$ letters. Subsequent works~\cite{DBLP:conf/esa/LoukidesP21,DBLP:journals/tkde/LoukidesPS23,DBLP:journals/pvldb/AyadLP23} propose to also construct a sparse suffix array for the reversed prefixes
ending at the positions in $B$, and conceptually link the two suffix arrays with a geometric data structure.
As opposed to~\cite{DBLP:journals/jda/ClaudeNPST12,DBLP:journals/spe/GrabowskiR17}, these approaches~\cite{DBLP:conf/esa/LoukidesP21,DBLP:journals/tkde/LoukidesPS23,DBLP:journals/pvldb/AyadLP23} thus offer query times with theoretical guarantees. An important practical limitation of these works is that they rely on sparse suffix sorting which is a rather undeveloped topic in practical terms~\cite{DBLP:conf/latin/AyadLPV24}.
From a theory perspective, the following is known.


\begin{theorem}[\cite{DBLP:journals/corr/abs-2407-11819}]\label{the:worst-case}
For any string $T=T[0\dd n)$ over an alphabet $\Sigma=[0,\sigma)$ with $\sigma=n^{\cO(1)}$ and any integer $\ell>0$, we can construct an index that occupies $\cO(n/\ell)$ extra space and reports all $\Count(P,T)$ occurrences of any pattern $P$ of length $|P|\geq \ell$ in $T$ in $\ctO(|P|+\Count(P,T))$ time. The index can be constructed in $\ctO(n)$ time and $\cO(n/\ell)$ working space.
\end{theorem}

The practical limitation of Theorem~\ref{the:worst-case} is that it relies on
an intricate sampling scheme and on geometric data structures
which are both unlikely to be efficient in practice~\cite{DBLP:journals/corr/abs-2407-11819}.


Another common characteristic of the aforementioned approaches 
is that \emph{they are not universal}. They enhance the text with specific data structures (typically, the sparse suffix array of the sampled suffixes and some geometric data structures) and so they also have a specific query algorithm. The main benefit of the approach we describe in this paper is that it can be used with (and improve) 
\emph{any} text indexing technique~\cite{DBLP:journals/siamcomp/ManberM93,10.1145/1082036.1082039,grossi2005compressed,10.1145/3375890,DBLP:conf/soda/KempaK23,DBLP:conf/focs/KempaK23}.

\subparagraph{Other Related Work.}
There also exists work~\cite{accelerated_fm_pfp} that attempts to accelerate indexing lookup by working in sketch space (in this case, using a prefix-free parse~\cite{Boucher_2019} of the text and pattern). This approach, however, builds an index over both the original and the sketched text, and has been explored only in the context of compressed indexes (i.e., the FM-index).




\section{A Universal Indexing Framework for Matching Long Patterns}\label{sec:framework}

\begin{figure*}[htbp]
    \centering
    \includegraphics[width=1\linewidth]{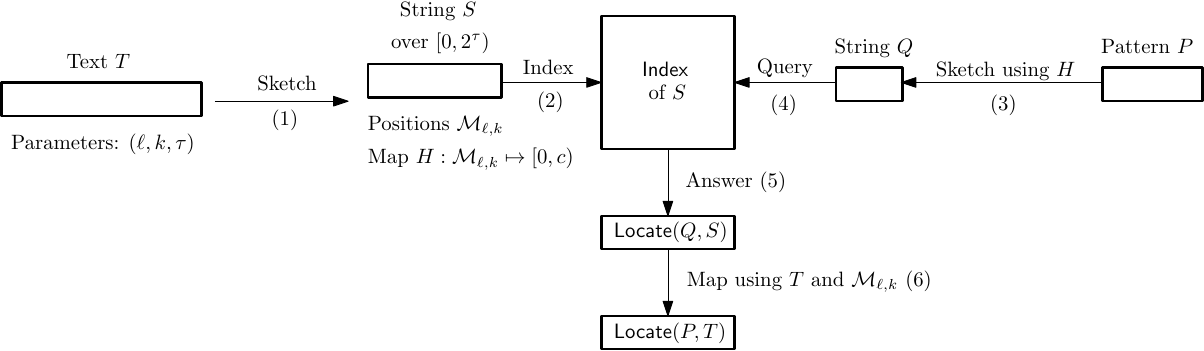}
    \caption{The \uindex framework. Steps (1) and (2) are to build the index. The steps (3)--(6) are to query with the framework. The sketching scheme in steps (1) and (3) must be the same.}
    \label{fig:framework}
\end{figure*}

In this section, we describe a universal indexing framework for a text $T$ of length $n$ --- referred to as the {\uindex} --- to retrieve all occurrences of a pattern $P$ of length $m\geq \ell$ in $T$.
Refer to \cref{fig:framework} for an overview of the proposed framework.

\subparagraph{Overview.}
The core idea of the {\uindex} is to sketch the text $T$.
We use random minimizers with parameters $k$ and $w:=\ell-k+1$ for sketching $T$,
though any type of locally consistent sketching mechanism may be used. We start by computing the sorted list $\Minimizers(T)$ of the minimizer \emph{positions}.
Let us set $z:=|\Minimizers(T)|$.
We also consider the sequence $M[0 \dd \Size)$ of the corresponding minimizer strings,
such that  $M[i] = T[p_i \dd p_i+k)$ for any position $p_i \in \Minimizers(T)$.
Let $c$ be the number of \textit{distinct} minimizers in $M$.
We can then identify each minimizer $v\in M$ with a unique identifier (or ``ID'' for short, in the following) in $[0,c)$ using a map $H : \Sigma^k \to [0,c)$.
The sketch $S[0 \dd \Size)=\Sketch(T)$ is the sequence of IDs $S[i]:=H(M[i])$ for all $i \in [0, \Size)$, which is encoded in a suitable alphabet.

Two remarks about the map $H$ are in order. When $k$ is small enough to have $\sigma^k \ll
n/\ell$, then most $k$-mers are likely to be minimizers and the map $H$ can thus be
completely omitted. In what follows, we assume the case when $H$ exists.   
When $k$ is large, on the other hand,
storing each minimizer in $M$ and the map $H$ could take a lot of space, e.g., $\cO(c (k\log_2(\sigma)+\log_2(c)))$ bits.
We can reduce the $k\log_2(\sigma)$ term of this space usage by first hashing the minimizers using, e.g., a rolling hash function (see \cref{sec:background}),
and only storing the mapping from minimizer hashes to their IDs.
This reduces the space to $\cO(c(q+\log_2(c)))$ bits, where $q$ is the number of bits used for each hash code, provided that $q < k\log_2(\sigma)$.


\subparagraph{Constructing the Index.}
Let $\Sigma'=[0,2^{\tau})$ denote the integer alphabet that we choose to encode $S$, for some input parameter $\tau$.
Further, let $\INDEX$ denote the chosen indexing data structure that we apply on $S$.
Namely, we construct the $\INDEX$ of $S$, over the alphabet $\Sigma'$, with $\tau=\log_2|\Sigma'|$. 
Note the purpose of setting the value of $\tau$: it lets the user control the size of the alphabet we choose to encode $S$ as something that lies in $[2,n]$.
Thus, we interpret each $\lceil\log_2(c)\rceil$-bit ID $S[i]$ as a sequence of $b:= \lceil \lceil\log_2(c)\rceil / \tau \rceil$ $\tau$-bit integers\footnote{In the unlikely event of $\Size\lceil\lceil\log_2(c)\rceil / \tau\rceil > n$, we can either increase $\tau$ to have $z\lceil\lceil\log_2(c)\rceil / \tau\rceil \leq n$ or simply set $S:=T$.}.
This is a useful feature because some compressed full-text indexes, like the FM-index~\cite{10.1145/1082036.1082039} or the $r$-index~\cite{10.1145/3375890}, take advantage of the repetitiveness of the text $T$ to improve its compression.

\subparagraph{Implicit Sketched Text.}
Note that the $\INDEX$ of $S$ may or may not require storing the sketched text $S$ itself. For example, the FM-index is a \emph{self-index} and replaces $S$ with its compressed form. On the other hand, the suffix array is not a self-index and \emph{does} require $S$. 
In the latter case, 
we can either store
$S$ explicitly, or we can reconstruct $S$ \emph{on-the-fly} as needed using only $T$, $H$, and $\Minimizers(T)$.


To conclude, our framework assumes read-only random access to $T$, takes
parameters $\ell$, $k$, and $\tau$ as input, and constructs an index on top of
$T$ that consists only of the minimizer positions $\Minimizers(T)$ (encoded
using Elias-Fano~\cite{Elias74,Fano71}), the minimizer-ID map $H$, and the $\INDEX$ of $S$ over a $\tau$-bit alphabet.

\subparagraph{Querying.}
We now describe how to compute the set $L=\Locate(P,T)$, given a query pattern
$P[0\dd m)$ that is sufficiently long (i.e., $m\geq \ell$).

First, $P$ is sketched similarly to the text $T$, obtaining a string $Q=\Sketch(P)$.
Specifically, its minimizer positions
$\Minimizers(P)$ are found. Since the pattern has length $m \geq \ell$,
it has at least one minimizer, and we indicate with $\alpha$ and $\beta$ the position
of the first and last minimizer of $P$, respectively. If one of the minimizers $P[p_i\dd p_i+k)$ of $P$, for $p_i\in \Minimizers(P)$, 
does not occur in the text $T$ and hence is not assigned
an ID by $H$, this directly implies that $P$ does not occur in $T$.
Otherwise, the list of corresponding IDs is determined as
$H(P[p_i \dd p_i+k))$, 
for all $p_i\in \Minimizers(P)$, and this is encoded into the
sketched query string $Q$ using $b$ $\tau$-bit integers per minimizer.

\begin{figure*}[t]
    \centering
    \includegraphics[width=1\linewidth]{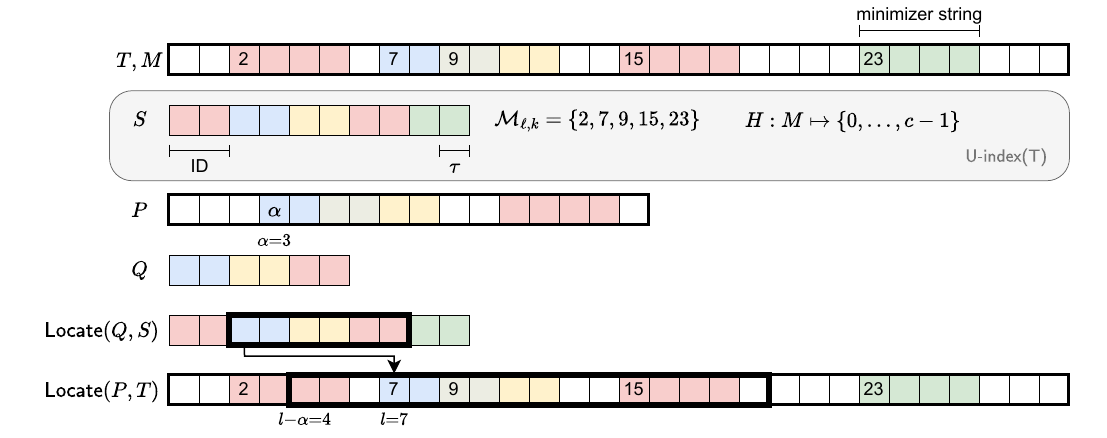}
    \caption{An illustration of the \uindex of a text $T$, along with a query example. First, the minimizers $M$ of $T$
      are found, here of length $k=4$ characters, with two of them overlapping (those starting at positions 7 and 9).
      The minimizer positions $\Minimizers$ are stored with
      Elias-Fano coding. Minimizers are hashed via $H$ to shorter IDs. These are
      padded to the next multiple of $\tau$.
      An index is then built on the sketch $S$.
      To locate a pattern $P$,
      its minimizers are found and the sketch $Q$ of corresponding IDs is constructed. Then
      $Q$ is located in $S$, which here gives a single match. The first
      minimizer of the match in $Q$ is located in $T$ at position $l$ via $\Minimizers$.
      Lastly, the candidate match is verified starting at position $l-\alpha$ in
      $T$.}
    \label{fig:example}
\end{figure*}

We locate $Q$ in $S$ using the $\INDEX$ of $S$. Let $L=\varnothing$ be the initially empty list
of occurrences.
For every position
$p \in L' = \Locate(S, Q)$, we first check whether $p \equiv 0 \pmod b$.
If not, the candidate match is a false positive caused by the reduction of alphabet
size. Otherwise, we retrieve the position $l:=\Minimizers(T)[p/b]$ and verify whether $T[l-\alpha \dd l-\alpha + m)=P$ in $\cO(m)$ time.
If so, position $(l-\alpha)$ is added to $L$.
\cref{fig:example} illustrates an example.

\subparagraph{Theoretical Guarantees.}
We now explain how to verify an occurrence at position $p \in L'$ in $\cO(1)$ time using $\cO(\Size)$ space.
Let the occurrence be $S[p \dd q]$,
where $q=p+|Q|-1$. 

For faster querying in theory or for very long patterns in practice, we also store an array $F$ of fingerprints, where $F[i] = \phi(T[0 \dd p_i])$, for all $p_i\in \Minimizers(T)$, and $\phi$ is the rolling hash function; see \cref{sec:background}. 
This array can be constructed in $\cO(n)$ time and has size $\cO(\Size)$.
Recall that $\ell$ is the lower bound on the length of $P$.
Let $\mathcal{X} = \{T[p_i\dd p_i+\ell) \mid p_i \in \Minimizers(T)\}$ and $\mathcal{X}^R = \{(T[p_i-\ell+1\dd p_i])^R \mid p_i \in \Minimizers(T)\}$, where $s^R$ denotes the reverse of the string $s$.
We construct the tries $\mathcal{T} := \TRIE(\mathcal{X})$ and $\mathcal{T}^R := \TRIE(\mathcal{X}^R)$.
We label the leaf nodes representing the string $s=T[p_i\dd p_i+\ell)$ and $s^R=(T[p_i-\ell+1\dd p_i])^R$ in both tries by the set
$X_s = \{ p_i \mid T[p_i \dd p_i+\ell)=s \wedge p_i \in \Minimizers \}$. 
Each leaf node is also assigned a \emph{lex-rank} that is obtained via an in-order DFS traversal of the trie.
We also implement an inverse function that takes $p_i$ as input and returns the lex-rank of the leaf node that represents $s=T[p_i\dd p_i+\ell)$ in $\mathcal{T}$.
We implement the analogous inverse function for $\mathcal{T}^R$.
Each branching node $u$ in $\mathcal{T}$ stores an interval whose left and right endpoints are the lex-rank of the leftmost and rightmost leaf node, respectively,
in the subtree rooted at $u$. This information is also computed via a DFS traversal.
We store the analogous information for the branching nodes in $\mathcal{T}^R$.
Since $\mathcal{T}$ and $\mathcal{T}^R$ are compacted and $\sum |X_s| = \Size$, it follows that the tries and the inverse functions take $\cO(\Size)$ space. Furthermore, they can be constructed in $\cO(n)$ time~\cite{DBLP:journals/jea/Charalampopoulos20}.

\begin{figure}
    \centering
    \includegraphics[width=1\linewidth]{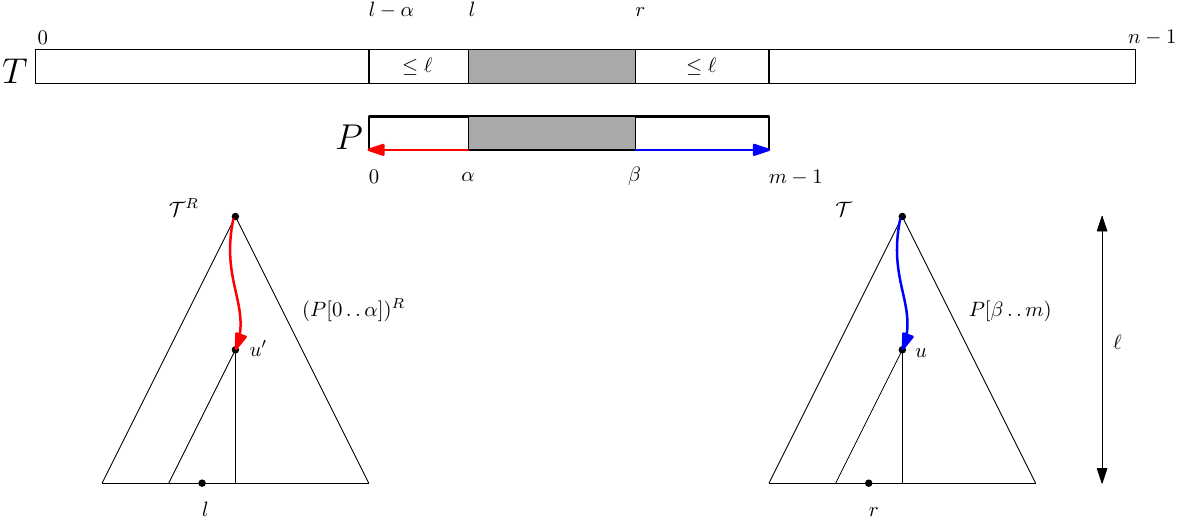}
    \caption{The $\cO(1)$-time verification algorithm for occurrence $l-\alpha$. After spelling the fragments of $P$ in the two tries \emph{once}, we check if the fragments in gray match using fingerprints in $\cO(1)$ time; if so, we check if the corresponding leaf nodes are \emph{both} located in the induced subtrees in $\cO(1)$ time.}
    \label{fig:verification}
\end{figure}

Let us explain how these additional structures can help us verify an occurrence
$S[p\dd q]$ of $Q$ in $S$ in $\cO(1)$ time; see \cref{fig:verification}. Let $l:=\Minimizers(T)[p/b]$ and $r:=\Minimizers(T)[q/b]$.
Using the vector $F$,
we compute $\phi(T[l+1\dd r])$ in $\cO(1)$ time by Fact~\ref{fact:KR}, because we have
$F[p]=\phi(T[0\dd l])$ and $F[q]=\phi(T[0\dd r])$.
We also compute the
fingerprint $\phi(P[\alpha+1 \dd \beta])$
once in $\cO(m)$ time and compare the two fingerprints in $\cO(1)$ time\footnote{If $|\Minimizers(P)|=1$, then we always return a positive answer for this comparison.}. If they are not equal, then $(l-\alpha)$ is not a valid occurrence. If they are equal, we need to check $P[0\dd \alpha]$ and $P[\beta\dd m)$.
The remaining letters on each edge cannot be more than $\ell$ (by the density of the minimizers mechanism), and so the verification would cost $\cO(\ell)$ time if we did it by letter comparisons. We can verify the edges in $\cO(1)$ time using tries; see \cref{fig:verification}. In a preprocessing step, we spell $P[\beta\dd m)$ in $\mathcal{T}$ arriving at node $u$; and we spell $(P[0\dd \alpha])^R$ in $\mathcal{T}^R$ arriving at node $u'$. This takes $\cO(\ell)$ time. We can then check whether $r$ is a leaf node in the subtree induced in $\mathcal{T}$ using the inverse function and the interval stored in $u$. We can also check whether $l$ is a leaf node in the subtree induced in $\mathcal{T}^R$ using the inverse function and the interval stored in $u'$. This takes $\cO(1)$ time per pair $(l,r)$.
We then have that $(l-\alpha)$ is a valid occurrence if and only if both leaf nodes are located in the induced subtrees. 

\return
We have thus arrived at the following result.


\begin{theorem}(Universal framework)\label{the:framework}
Let $T$ be a string of length $n$ over alphabet
  $\Sigma=[0,\sigma)$. Let $t(n,\sigma)$, $s(n,\sigma)$, and $q(m,n,\sigma)$ be,
    respectively, the time complexity to construct \INDEX, the size of \INDEX in
    machine words, and the query time of \INDEX to report all the occurrences of
    a pattern of length $m$ in $T$.
    Furthermore, let $\Size:=|\Minimizers(T)|$ be the number of minimizers of $T$,
    for some parameters $\ell,k$, and let $S$ be the string obtained from $T$
    using the framework with a parameter $\tau$ chosen from $[1, \log_2(n)]$. Then, in
    $\cO(n+t(\Size\lceil\log \Size/\tau\rceil,2^\tau))$ time, we can construct an
    index of $\cO(\Size+s(\Size\lceil\log \Size/\tau\rceil,2^\tau))$ size,
    supporting $\Locate(P,T)$ queries for a pattern $P$ of length $m\geq \ell$ in $\cO(m+q(|Q|,\Size\lceil\log \Size/\tau\rceil,2^\tau)+\Count(Q,S))$ time, where $Q$ is the string obtained from $P$ using the framework with parameters $\ell,k,\tau$. 
\end{theorem}


\subparagraph{Example.} Let us now consider a practical instantiation of \cref{the:framework}. Let \INDEX be the \emph{suffix array}~\cite{DBLP:journals/siamcomp/ManberM93} enhanced with the \emph{longest
common prefix} (LCP) array~\cite{DBLP:conf/cpm/KasaiLAAP01}. We choose
$\tau:=\lceil\log_2 n\rceil$ because the suffix array can be constructed in $t(n,2^\tau)=\cO(n)$ time, for any integer alphabet of size $2^\tau\leq n$,  and it has size $s(n,2^\tau)=\cO(n)$~\cite{DBLP:journals/jacm/KarkkainenSB06}.
Given the suffix array, the LCP array can be constructed in $\cO(n)$ time~\cite{DBLP:conf/cpm/KasaiLAAP01}. By applying \cref{the:framework}, we construct a string $S$ of length $\Size\leq n$
over the alphabet $[0,n)$. Thus, we will construct our index
  of $\cO(\Size+s(\Size,n))=\cO(\Size)$ size in $\cO(n+t(\Size,n))=\cO(n)$ time. Note that, by
  using minimizers~\cite{DBLP:journals/bioinformatics/RobertsHHMY04,DBLP:conf/sigmod/SchleimerWA03}, $\Size$ can be much smaller
than $n$ in practice, for a sufficiently large value of $\ell$. Specifically, we have $\Size = nd$ where $d \geq 1/(\ell-k+1)$ is the \textit{density} of the specific minimizer scheme used.
For querying, we have $q(m,n,2^\tau)=\cO(m+\log n + \Count(P,T))$ when LCP information is used~\cite{DBLP:journals/siamcomp/ManberM93}.
Thus, our query time is $\cO(m+\log \Size+\Count(Q,S))$
because $m\geq |Q|$ and $\Count(Q,S)\geq \Count(P,T)$. We stress that even if $\Count(Q,S)\geq \Count(P,T)$, we also have $\log \Size \leq \log n$, and so beyond space savings, the resulting index can also be competitive or faster in querying.

\section{Experiments}\label{sec:experiments}

We implemented the {\uindex} framework in the Rust programming language.
In \cref{sec:setup}, we present the setup of the experiments that we conducted to assess the efficiency of our implementation.
In \cref{sec:evaluation}, we present the results of these experiments.
In \cref{sec:application}, we present an application of our framework in mapping long reads onto a reference genome.

Our software resources are open-source and can be found at \url{https://github.com/u-index/u-index-rs}.

\subparagraph{Implementation Details.}
In our implementation, we verify each candidate occurrence using a linear scan of $P$ in
$T$, hence without using any trie data structure. Even if this solution costs
$\cO(m)$ time instead of the $\cO(1)$-time verification claimed by
Theorem~\ref{the:framework}, this is likely to be faster in practice because (as it is well known) traversing tries is not cache-efficient.

\subsection{Setup}\label{sec:setup}

\subparagraph{Hardware and Software.}
All experiments were run on an Intel Core
i7-10750H running at a fixed frequency of 2.6 GHz with hyperthreading disabled
and cache sizes of 32 KiB (L1), 256 KiB (L2), and 12 MiB (shared L3). Code was
compiled using \verb|rustc| 1.85.0-nightly and \verb|GCC|
14.2.1 on Arch Linux.

\begin{figure*}[t]
\centering
\includegraphics[width=\textwidth]{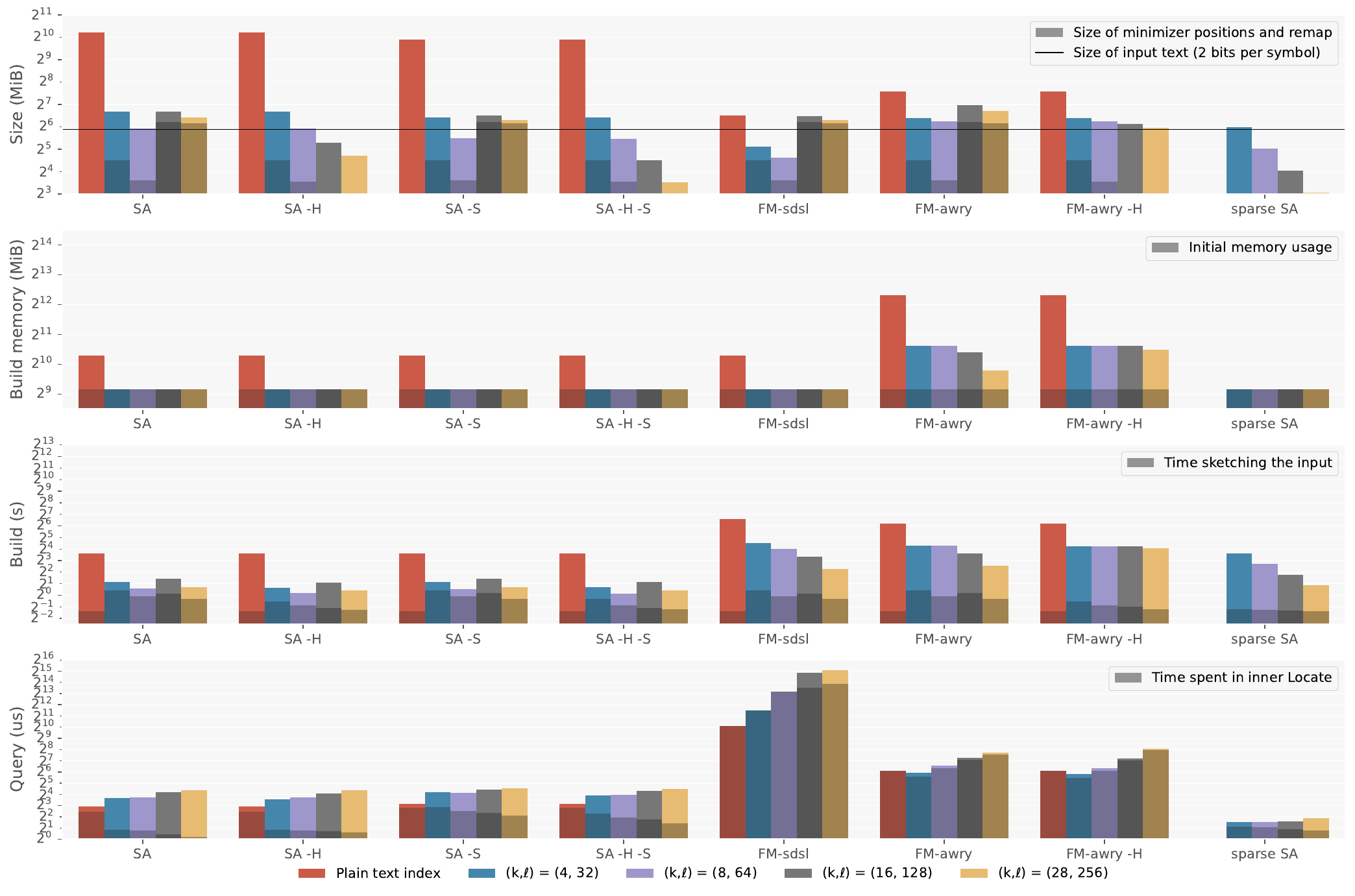}%
\caption{
Results on the $59$ MiB of the human chromosome 1. For each data structure --- except the {\sindex} --- we compare its performance when constructed on the plain input text (in red, left column of each group) versus when used with the {\uindex} (remaining
  colors and columns), for increasing values of $k$ and $\ell$.
  Indexes marked with \texttt{-H} (read, ``minus \texttt{H}'') use minimizers themselves as IDs, \textit{without} the map $H$.
  Similarly,
  the indexes marked with \texttt{-S} omit storing the sketched input text
  $S$ and instead reconstruct it via the minimizer positions $\Minimizers(T)$
  and $T$ itself.
  The {\sindex} is only shown with sampling (no red column) because it is otherwise equivalent
  to {{\sa}}.
  The top plot shows the space usage (size) of the final data structure in MiB, with the space for minimizer positions
  and the map $H$ shaded, and the black line indicating the space occupied by the
  2-bit packed input text. The second row shows the maximum memory usage (resident set size) during the
  construction, where the shaded area is the memory usage \emph{before} construction.
  The third row shows the construction time (in seconds), with the time for sketching the input shaded.
  The bottom plot shows the query time (in average $\mu$s per {\Locate} query), with the time for searching in the
  inner index shaded.
  }
\label{fig:plot-v2}
\end{figure*}

\subparagraph{Datasets.}
We use three textual datasets of different nature and alphabet size:
(1) chromosome 1 of
CHM13v2.0\footnote{\url{https://www.ncbi.nlm.nih.gov/nuccore/NC_060925.1}},
which contains repetitive regions and consists of 248 million characters over the DNA alphabet ($\sigma=4$),
thus 59 MiB when each character is coded using 2 bits;
(2) the 200 MiB protein sequences available from the Pizza \& Chili site\footnote{\url{https://pizzachili.dcc.uchile.cl/texts/protein}} ($\sigma=27$), or 125 MiB when each character is coded using 5 bits;
and (3) the 200 MiB English collection, also available from the Pizza \& Chili site\footnote{\url{https://pizzachili.dcc.uchile.cl/texts/nlang}} ($\sigma=239$).
In the following, we discuss experimental results referring to \cref{fig:plot-v2} (on page \pageref{fig:plot-v2}) for the DNA alphabet. The results for proteins and English have very similar shapes and are deferred to \cref{sec:app} due to space constraints.

\subparagraph{Queries.} For each dataset, we test $10^5$ \emph{positive} queries that
are uniformly sampled from the text. For DNA, queries consist of $512$ characters,
while for the protein and English datasets, we use queries of $128$ characters.

\subparagraph{Tested Indexes.}
We compare the suffix array and the FM-index, indicated with {\sa} and {\fmindex} in our results, against their
{\uindex} variants, with parameters $(k,\ell)\in \{(4,32), (8,64), (16,128), (28,256)\}$.
For the suffix array construction, we use
\method{libsais}\footnote{\url{https://github.com/IlyaGrebnov/libsais}}~\cite{nong2010two,karkkainen2009permuted}.
For the {\fmindex}, we use the implementations in SDSL-lite
(SDSL-v2)~\cite{gog2014theory} and in AWRY~\cite{anderson2021optimized}.

For each index, we use the smallest $\tau\geq \lceil\log_2 c \rceil$ that is supported by
the index. In practice, this means that for the suffix array and SDSL {\fmindex}
we use $\lceil \log_2 c /8\rceil$ bytes to represent each ID, so that each ID is
a single character. This way, these indexes get built on exactly $z=|\Minimizers(T)|$ characters.
In practice, this is strictly better than using a smaller $\tau$ and building an
index on $2z$ or more characters.
The AWRY {\fmindex} does not support generic alphabets, and thus we had to consistently
use $\tau=2$ to encode the IDs as DNA bases. We note that the AWRY {\fmindex} uses a
multi-threaded parallel construction algorithm, while all other methods are
single-threaded.
Further details can be found in \cref{sec:app-tau}.

Finally, we also compare against our own implementation of the \emph{sparse suffix array at minimizer
positions}~\cite{DBLP:journals/spe/GrabowskiR17}, that we call the {\sindex} index in our results.

\subsection{Results}\label{sec:evaluation}

\subparagraph{Suffix Array.}
In all cases, the {\uindex} variants take less space and are faster to
construct than the classic indexes. 
Both space and
construction time tend to be at least $8\times$ less.
The time spent searching the suffix array (bottom, shaded) goes down as the
suffix array becomes sparser, but the total query time goes up.
This is due to the increasing number of false positive matches in minimizer
space, starting at $10$ false positives per query for $(k,\ell)=(4,32)$ and going
up to $100$ for $(k,\ell)=(28,256)$. These are caused by highly repetitive regions
in the centromere~\cite{grady1992highly}.
Nevertheless, the query time is usually less than $2\times$ slower than the classic suffix
array, while the space and construction time are greatly reduced.

\subparagraph{Omitting $H$.}
When $k\in \{16,28\}$, most minimizers are unique, and the map $H$ mapping the minimizers
to IDs has size linear in their number. This can be seen in that the shaded area in
the rightmost two columns of the top-left group is accountable for most of the
space used by the index at over $2^6=64$ MiB, while the input only takes 59 MiB when encoded using $2$ bits per character.
Looking at {\sa} \texttt{-H}, which omits $H$,
we see that the index is much smaller for larger $k$, while the query time is
unaffected. The construction time does increase though, since the alphabet is larger and
hence requires more memory. The high spike for $k=16$ seems to be caused by
{\sa} (\method{libsais}) being particularly slow for 32-bit integers.

\subparagraph{Implicit Sketched Text.}
We can also omit the sketched text $S$ and
instead reconstruct it on-the-fly, as explained in \cref{sec:framework}. This saves significant space when $k$ is
large (where also $H$ can be omitted altogether). The construction time is not affected since $S$ is discarded after the construction. The time spent in searching the suffix array
does increase significantly though, up to $4\times$ (shaded), but as most of the time
is spent verifying potential matches, the total query time only goes up slightly.

\subparagraph{{\fmindex}.}
The SDSL implementation takes around 90 MiB for the plain input text, and goes significantly below this when using the {\uindex}\footnote{For some technical reason, SDSL does not accept NULL bytes, hence minimizers must necessarily be mapped by $H$ in the range $[1,c+1)$. Thus, SDSL takes a
significant amount of space for $k \geq 16$.}.
The construction time improves as well, almost
$2\times$ each time we double $\ell$ because the number of sampled positions nearly halves.

A main drawback of the SDSL implementation is its significantly larger query times compared to all
other methods, starting at $32\times$ slower for the plain index and increasing up to $1000\times$ slower for $\ell=256$.
We suspect this is due to the inherent complexity in the wavelet tree data structures used to represent the Burrows-Wheeler transform of the text~\cite{burrows1994block}, that has increasingly more levels as $k$ increases and hence the number of bits $\tau$ in each $k$-mer ID grows.
This makes SDSL an impractical choice in this scenario or, more in general, for applications where the final size is not a bottleneck but fast queries are the primary concern.

AWRY uses around twice as much space as SDSL, and has a similar construction
time, likely because both are limited by the internal suffix array construction.
On the other hand, query times for AWRY are significantly faster than SDSL because AWRY is optimized for the DNA alphabet, and the {\uindex} version
with $k=4$ is slightly faster than AWRY on the plain text. However, as $k$ and
$\ell$ increase, both SDSL and AWRY slow down by roughly a factor of 2 overall.
For AWRY we can also omit $H$ and this almost halves the size for large $k$, but negates the previously seen speedup in construction time since $S$ is significantly larger.

\subparagraph{Sparse SA.}
Lastly, the {\sindex} index takes strictly less space than the {\uindex} variant of the suffix
array, since it stores strictly less information: only one permutation of the
minimizer positions in the original text.
It is also significantly faster to query, since no sketching is needed.
Additionally, it has much fewer false positives, since comparisons are made in
the plain input space rather than in sketch space. The construction time is worse
than {\sa}, however, since there is no available linear-time implementation for
sparse suffix array construction\footnote{Indeed, our implementation is very simple and relies on
sorting the minimizer positions using the Rust
\texttt{sort\_by} function that returns the suffix associated with each
position as needed.}.
Nevertheless, this is still faster than constructing an {\fmindex}.


\subsection{An Example Application}\label{sec:application}

To show a concrete application of our framework, we show that {\uindex} can be
used for \textit{long-read mapping}.
This is the problem of aligning long DNA or RNA sequencing reads (e.g., of length more than 1,000 base pairs) to a
reference genome.
Although long reads offer significant advantages over short reads in many
crucial tasks in bioinformatics,
such as genome assembly or structural variant detection,
their alignment is computationally costly.

\subparagraph{Setup.}
We run the following experiment.
As input, we take the full human genome (CHM13v2.0) and
a prefix of 450 PacBio HiFi long reads from the HG002-rep1 dataset\footnote{Downloaded from
\url{https://downloads.pacbcloud.com/public/revio/2022Q4/HG002-rep1/}.}.
These reads are approximately 99.9\% accurate and have an average length of
16kbp.

We partition (chunk) each read into patterns of length 256, and search each
of these patterns in our index. Short leftover suffixes are ignored.
Ideally, each read has then at least one pattern that
exactly matches the text, which can then be used to anchor an alignment.

\subparagraph{Results.}
We build the {\uindex} on top of the \method{libsais} suffix array in the
configuration where the sketched text $S$ and map $H$ are both stored.
We use $k=8$ and $\ell=128$, so that at least 2 and on average 4 minimizers are
sampled from every pattern.
This results in 53M minimizers, and the entire {\uindex} is built in 12 seconds.

Out of the 450 reads, 445 have at least one matching pattern, and in total,
14\ 824 of the 28\ 243 patterns match (52\%).
As observed before, an issue with DNA is that it contains many long repetitive
regions. In particular, there are 160 patterns that match around 3820 times each for a total of 611k matches, while the 28k remaining patterns only match 27k in total.
Worse, there are 721M mismatches, i.e., candidate matches in sketch space that
turn out not to be matches in the genome. Verifying these candidate matches
takes over 98\% of the time.

\subparagraph{Limiting Matches.}
Only very few (<200) patterns have more than 10 matches, and thus we stop
searching once we hit 10 matches. Further, most patterns have
relatively few mismatches, while a few patterns have a lot of mismatches. To
also avoid those negative effects, we generally only consider the first 100
matches in sketch space. We still match 445 of 450 reads, while the number
of matched patterns goes down to 13\ 717. On the other hand, the number of
mismatches is now 663k (23 per query) and the number of matches is 25k (0.9 per
query).

The result is a query time of 8.7$\mu$s per pattern or 550$\mu$s per read, of which 33\% is sketching the input, 18\% is locating the sketch, and the remaining 48\% is verification.
\section{Conclusions and Future Work}\label{sec:conclusions}

In this work, we introduced the {\uindex} --- a universal framework to enhance the performance of any off-the-shelf text index, provided that the patterns to match are sufficiently long.
This is achieved, in short, by sketching the text and using any desired index
for the sketched text.
Intuitively, this saves resources at building time and considerably reduces the
final index size, simply
because the sketched text is shorter than the input text.
Our experiments indeed confirm that the \uindex 
has excellent performance when used in combination with the suffix array, as
it significantly improves index size and construction while not slowing down queries too much. When paired with the FM-index, the savings are more modest but still significant.

The \textit{sparse} suffix array index by Grabowski and Raniszewski~\cite{DBLP:journals/spe/GrabowskiR17} remains a great solution in
this regime, having a smaller size and significantly faster queries than the
\uindex around suffix arrays (which have, however, faster construction time).
For example, albeit somewhat larger than the SDSL FM-index, the sparse suffix array is over $100\times$ faster to query.
However, the benefit of the {\uindex} framework lies in its universality and usability.
We remark that the primary objective of this work is to highlight
these two important properties.

We anticipate that the \uindex may be especially useful around the r-index~\cite{10.1145/3375890} when used on highly repetitive data, but we leave this as future work.
The sparse suffix array will \emph{by design} be unable to take advantage of the underlying repetitiveness. Hence, other than universality, another important virtue of our framework is that it preserves string similarity: for any two highly similar texts $T_1$ and $T_2$, it will be the case that $S_1=\Sketch(T_1)$ and $S_2=\Sketch(T_2)$ are also highly similar (e.g., assuming the sketches are based on minimizers).

In terms of theory, it would make sense to bound
$\Count(Q,S)$ as a function of $\Count(P,T)$ and the sketching parameters.
Such a function would bound the number of false positives.

\bibliographystyle{plain}
\bibliography{bibliography}

\begin{thebibliography}{10}

\bibitem{anderson2021optimized}
Tim Anderson and Travis~J Wheeler.
\newblock An optimized {FM}-index library for nucleotide and amino acid search.
\newblock {\em Algorithms for Molecular Biology}, 16(1):25, 2021.

\bibitem{u-index-arxiv-version}
Lorraine A.~K. Ayad, Gabriele Fici, Ragnar~Groot Koerkamp, Grigorios Loukides,
  Rob Patro, Giulio~Ermanno Pibiri, and Solon~P. Pissis.
\newblock U-index: A universal indexing framework for matching long patterns.
\newblock {\em CoRR}, abs/2502.14488, 2025.

\bibitem{DBLP:journals/pvldb/AyadLP23}
Lorraine A.~K. Ayad, Grigorios Loukides, and Solon~P. Pissis.
\newblock Text indexing for long patterns: Anchors are all you need.
\newblock {\em Proc. {VLDB} Endow.}, 16(9):2117--2131, 2023.

\bibitem{DBLP:journals/corr/abs-2407-11819}
Lorraine A.~K. Ayad, Grigorios Loukides, and Solon~P. Pissis.
\newblock Text indexing for long patterns using locally consistent anchors.
\newblock {\em CoRR}, abs/2407.11819, 2024.

\bibitem{DBLP:conf/latin/AyadLPV24}
Lorraine A.~K. Ayad, Grigorios Loukides, Solon~P. Pissis, and Hilde Verbeek.
\newblock Sparse suffix and {LCP} array: Simple, direct, small, and fast.
\newblock In Jos{\'{e}}~A. Soto and Andreas Wiese, editors, {\em {LATIN} 2024:
  Theoretical Informatics - 16th Latin American Symposium, Puerto Varas, Chile,
  March 18-22, 2024, Proceedings, Part {I}}, volume 14578 of {\em Lecture Notes
  in Computer Science}, pages 162--177. Springer, 2024.

\bibitem{Boucher_2019}
Christina Boucher, Travis Gagie, Alan Kuhnle, Ben Langmead, Giovanni Manzini,
  and Taher Mun.
\newblock {Prefix-free parsing for building big BWTs}.
\newblock {\em {Algorithms for Molecular Biology}}, 14(1), May 2019.

\bibitem{burrows1994block}
Michael Burrows and David Wheeler.
\newblock A block-sorting lossless data compression algorithm.
\newblock Technical report, Systems Research Center, 1994.

\bibitem{DBLP:journals/jea/Charalampopoulos20}
Panagiotis Charalampopoulos, Costas~S. Iliopoulos, Chang Liu, and Solon~P.
  Pissis.
\newblock Property suffix array with applications in indexing weighted
  sequences.
\newblock {\em {ACM} J. Exp. Algorithmics}, 25:1--16, 2020.

\bibitem{DBLP:journals/jda/ClaudeNPST12}
Francisco Claude, Gonzalo Navarro, Hannu Peltola, Leena Salmela, and Jorma
  Tarhio.
\newblock String matching with alphabet sampling.
\newblock {\em J. Discrete Algorithms}, 11:37--50, 2012.

\bibitem{DBLP:conf/icalp/DietzfelbingerGMP92}
Martin Dietzfelbinger, Joseph Gil, Yossi Matias, and Nicholas Pippenger.
\newblock Polynomial hash functions are reliable (extended abstract).
\newblock In Werner Kuich, editor, {\em Automata, Languages and Programming,
  19th International Colloquium, ICALP92, Vienna, Austria, July 13-17, 1992,
  Proceedings}, volume 623 of {\em Lecture Notes in Computer Science}, pages
  235--246. Springer, 1992.

\bibitem{Syncmers}
Robert Edgar.
\newblock Syncmers are more sensitive than minimizers for selecting conserved
  k‑mers in biological sequences.
\newblock {\em PeerJ}, 9(e10805):1755--1771, 2021.

\bibitem{Elias74}
Peter Elias.
\newblock Efficient storage and retrieval by content and address of static
  files.
\newblock {\em J. {ACM}}, 21(2):246--260, 1974.

\bibitem{Fano71}
Robert~Mario Fano.
\newblock On the number of bits required to implement an associative memory.
\newblock {\em Memorandum 61, Computer Structures Group, MIT}, 1971.

\bibitem{10.1145/1082036.1082039}
Paolo Ferragina and Giovanni Manzini.
\newblock Indexing compressed text.
\newblock {\em J. {ACM}}, 52(4):552--581, 2005.

\bibitem{10.1145/3375890}
Travis Gagie, Gonzalo Navarro, and Nicola Prezza.
\newblock Fully functional suffix trees and optimal text searching in
  {BWT}-runs bounded space.
\newblock {\em J. {ACM}}, 67(1):2:1--2:54, 2020.

\bibitem{gog2014theory}
Simon Gog, Timo Beller, Alistair Moffat, and Matthias Petri.
\newblock From theory to practice: Plug and play with succinct data structures.
\newblock In Joachim Gudmundsson and Jyrki Katajainen, editors, {\em
  Experimental Algorithms - 13th International Symposium, {SEA} 2014,
  Copenhagen, Denmark, June 29 - July 1, 2014. Proceedings}, volume 8504 of
  {\em Lecture Notes in Computer Science}, pages 326--337. Springer, 2014.

\bibitem{DBLP:journals/spe/GrabowskiR17}
Szymon Grabowski and Marcin Raniszewski.
\newblock Sampled suffix array with minimizers.
\newblock {\em Softw. Pract. Exp.}, 47(11):1755--1771, 2017.

\bibitem{grady1992highly}
Deborah~L Grady, Robert~L Ratliff, Donna~L Robinson, Erin~C McCanlies, Julianne
  Meyne, and Robert~K Moyzis.
\newblock Highly conserved repetitive {DNA} sequences are present at human
  centromeres.
\newblock {\em Proceedings of the National Academy of Sciences},
  89(5):1695--1699, 1992.

\bibitem{grootkoerkamp_et_al:LIPIcs.WABI.2024.11}
Ragnar Groot~Koerkamp and Giulio~Ermanno Pibiri.
\newblock {The mod-minimizer: A Simple and Efficient Sampling Algorithm for
  Long k-mers}.
\newblock In Solon~P. Pissis and Wing-Kin Sung, editors, {\em 24th
  International Workshop on Algorithms in Bioinformatics (WABI 2024)}, volume
  312 of {\em Leibniz International Proceedings in Informatics (LIPIcs)}, pages
  11:1--11:23, Dagstuhl, Germany, 2024. Schloss Dagstuhl -- Leibniz-Zentrum
  f{\"u}r Informatik.

\bibitem{grossi2005compressed}
Roberto Grossi and Jeffrey~Scott Vitter.
\newblock Compressed suffix arrays and suffix trees with applications to text
  indexing and string matching.
\newblock {\em {SIAM} J. Comput.}, 35(2):378--407, 2005.

\bibitem{gusfield1997algorithms}
Dan Gusfield.
\newblock {\em Algorithms on Strings, Trees, and Sequences - Computer Science
  and Computational Biology}.
\newblock Cambridge University Press, 1997.

\bibitem{accelerated_fm_pfp}
Aaron Hong, Marco Oliva, Dominik K{\"{o}}ppl, Hideo Bannai, Christina Boucher,
  and Travis Gagie.
\newblock Acceleration of fm-index queries through prefix-free parsing.
\newblock In Djamal Belazzougui and A{\"{\i}}da Ouangraoua, editors, {\em 23rd
  International Workshop on Algorithms in Bioinformatics, {WABI} 2023,
  September 4-6, 2023, Houston, TX, {USA}}, volume 273 of {\em LIPIcs}, pages
  13:1--13:16. Schloss Dagstuhl - Leibniz-Zentrum f{\"{u}}r Informatik, 2023.

\bibitem{karkkainen2009permuted}
Juha K{\"{a}}rkk{\"{a}}inen, Giovanni Manzini, and Simon~J. Puglisi.
\newblock Permuted longest-common-prefix array.
\newblock In Gregory Kucherov and Esko Ukkonen, editors, {\em Combinatorial
  Pattern Matching, 20th Annual Symposium, {CPM} 2009, Lille, France, June
  22-24, 2009, Proceedings}, volume 5577 of {\em Lecture Notes in Computer
  Science}, pages 181--192. Springer, 2009.

\bibitem{DBLP:journals/jacm/KarkkainenSB06}
Juha K{\"{a}}rkk{\"{a}}inen, Peter Sanders, and Stefan Burkhardt.
\newblock Linear work suffix array construction.
\newblock {\em J. {ACM}}, 53(6):918--936, 2006.

\bibitem{DBLP:journals/ibmrd/KarpR87}
Richard~M. Karp and Michael~O. Rabin.
\newblock Efficient randomized pattern-matching algorithms.
\newblock {\em {IBM} J. Res. Dev.}, 31(2):249--260, 1987.

\bibitem{DBLP:conf/cpm/KasaiLAAP01}
Toru Kasai, Gunho Lee, Hiroki Arimura, Setsuo Arikawa, and Kunsoo Park.
\newblock Linear-time longest-common-prefix computation in suffix arrays and
  its applications.
\newblock In Amihood Amir and Gad~M. Landau, editors, {\em Combinatorial
  Pattern Matching, 12th Annual Symposium, {CPM} 2001 Jerusalem, Israel, July
  1-4, 2001 Proceedings}, volume 2089 of {\em Lecture Notes in Computer
  Science}, pages 181--192. Springer, 2001.

\bibitem{DBLP:conf/soda/KempaK23}
Dominik Kempa and Tomasz Kociumaka.
\newblock Breaking the {{$\mathcal{O}$}(\emph{n})}-barrier in the construction
  of compressed suffix arrays and suffix trees.
\newblock In Nikhil Bansal and Viswanath Nagarajan, editors, {\em Proceedings
  of the 2023 {ACM-SIAM} Symposium on Discrete Algorithms, {SODA} 2023,
  Florence, Italy, January 22-25, 2023}, pages 5122--5202. {SIAM}, 2023.

\bibitem{DBLP:conf/focs/KempaK23}
Dominik Kempa and Tomasz Kociumaka.
\newblock Collapsing the hierarchy of compressed data structures: Suffix arrays
  in optimal compressed space.
\newblock In {\em 64th {IEEE} Annual Symposium on Foundations of Computer
  Science, {FOCS} 2023, Santa Cruz, CA, USA, November 6-9, 2023}, pages
  1877--1886. {IEEE}, 2023.

\bibitem{DBLP:conf/esa/LoukidesP21}
Grigorios Loukides and Solon~P. Pissis.
\newblock Bidirectional string anchors: {A} new string sampling mechanism.
\newblock In Petra Mutzel, Rasmus Pagh, and Grzegorz Herman, editors, {\em 29th
  Annual European Symposium on Algorithms, {ESA} 2021, September 6-8, 2021,
  Lisbon, Portugal (Virtual Conference)}, volume 204 of {\em LIPIcs}, pages
  64:1--64:21. Schloss Dagstuhl - Leibniz-Zentrum f{\"{u}}r Informatik, 2021.

\bibitem{DBLP:journals/tkde/LoukidesPS23}
Grigorios Loukides, Solon~P. Pissis, and Michelle Sweering.
\newblock Bidirectional string anchors for improved text indexing and top-{$K$}
  similarity search.
\newblock {\em {IEEE} Trans. Knowl. Data Eng.}, 35(11):11093--11111, 2023.

\bibitem{DBLP:journals/siamcomp/ManberM93}
Udi Manber and Eugene~W. Myers.
\newblock Suffix arrays: {A} new method for on-line string searches.
\newblock {\em {SIAM} J. Comput.}, 22(5):935--948, 1993.

\bibitem{navarro2007compressed}
Gonzalo Navarro and Veli M{\"{a}}kinen.
\newblock Compressed full-text indexes.
\newblock {\em {ACM} Comput. Surv.}, 39(1):2, 2007.

\bibitem{nong2010two}
Ge~Nong, Sen Zhang, and Wai~Hong Chan.
\newblock Two efficient algorithms for linear time suffix array construction.
\newblock {\em {IEEE} Trans. Computers}, 60(10):1471--1484, 2011.

\bibitem{DBLP:journals/bioinformatics/RobertsHHMY04}
Michael Roberts, Wayne Hayes, Brian~R. Hunt, Stephen~M. Mount, and James~A.
  Yorke.
\newblock Reducing storage requirements for biological sequence comparison.
\newblock {\em Bioinform.}, 20(18):3363--3369, 2004.

\bibitem{DBLP:conf/sigmod/SchleimerWA03}
Saul Schleimer, Daniel~Shawcross Wilkerson, and Alexander Aiken.
\newblock Winnowing: Local algorithms for document fingerprinting.
\newblock In Alon~Y. Halevy, Zachary~G. Ives, and AnHai Doan, editors, {\em
  Proceedings of the 2003 {ACM} {SIGMOD} International Conference on Management
  of Data, San Diego, California, USA, June 9-12, 2003}, pages 76--85. {ACM},
  2003.

\bibitem{zheng2020improved}
Hongyu Zheng, Carl Kingsford, and Guillaume Mar{\c{c}}ais.
\newblock Improved design and analysis of practical minimizers.
\newblock {\em Bioinform.}, 36(Supplement-1):i119--i127, 2020.

\end{thebibliography}

\newpage
\appendix
\section{Further Details on the Tested Indexes}\label{sec:app-tau}

In this section, we explain in more detail how each of the indexes was used.

\subparagraph{The \method{libsais} Suffix Array.}
For the plain input, including for DNA, we use a $\tau=8$ byte encoding.
For the sketched text $S$, this depends on $c$. When $\log_2(c)\leq 8$, we use one-byte
encoding and call the \texttt{libsais} function. When $\log_2(c)\leq 16$, we use
two-byte encoding and call \texttt{libsais16}. For larger $c$, we first remap
all IDs to values starting at $0$, as recommended by the library authors, and then call the function \texttt{libsais\_int} on 32-bit
input values.

\subparagraph{The SDSL-lite FM-index.}
For the plain text, we use the Huffman-shaped wavelet tree with default parameters, i.e., the class
\verb|csa_wt<wt_huff<rrr_vector<63>>,32,64>|.
For the {\uindex} counterpart, we instead use
\verb|csa_wt<wt_int<rrr_vector<63>>,32,64>|, a wavelet tree over a variable-width integer alphabet.
Both these indexes use a sampling factor of
$32$ to sample the suffix array. Also, for both versions we remap
$k$-mers to IDs starting at $1$ instead of $0$, since SDSL does not support input values of $0$.
For this reason, we do not have a \texttt{-H} no-remap variant of the SDSL {\fmindex}.

\subparagraph{The AWRY FM-index.}
The AWRY {\fmindex} only supports DNA and protein alphabets. For consistency, we
only use the DNA version. This means that we consistently use $\tau=2$. Thus,
IDs with $\log_2(c)$ bits are encoded into $\lceil\log_2(c)\rceil$ DNA characters,
which are passed to AWRY as 8-bit \texttt{ACGT} characters. Similarly, plain
text protein and English input is encoded into 4 underlying characters.
This index also uses a suffix array sampling factor of $32$.

\subparagraph{The \sindex.}
The \sindex consists of an array of 32-bit integers indicating text positions.
It is constructed using the Rust standard library
\verb|Vec<u32>::sort_unstable_by_key| function that compares text indices
by comparing the corresponding suffixes.

\clearpage
\section{Experimental Results on Proteins and English Datasets}\label{sec:app}

\begin{figure*}[h]
\centering
\includegraphics[width=\textwidth]{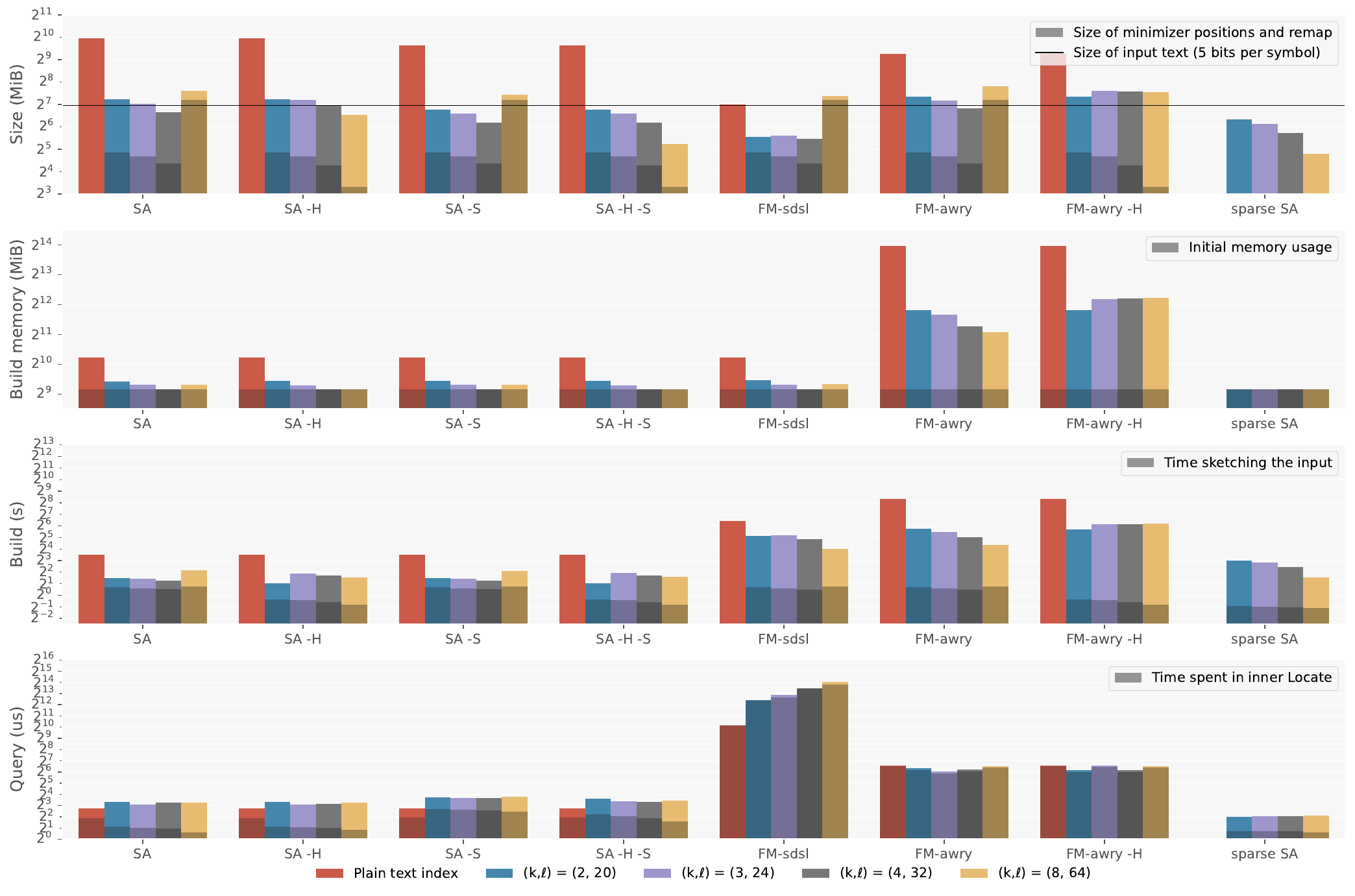}%
\caption{Results on $200$ MiB of protein sequences.
Refer to the caption of \cref{fig:plot-v2}.
}
\label{fig:plot-proteins}
\end{figure*}

\begin{figure*}[h]
\centering
\includegraphics[width=\textwidth]{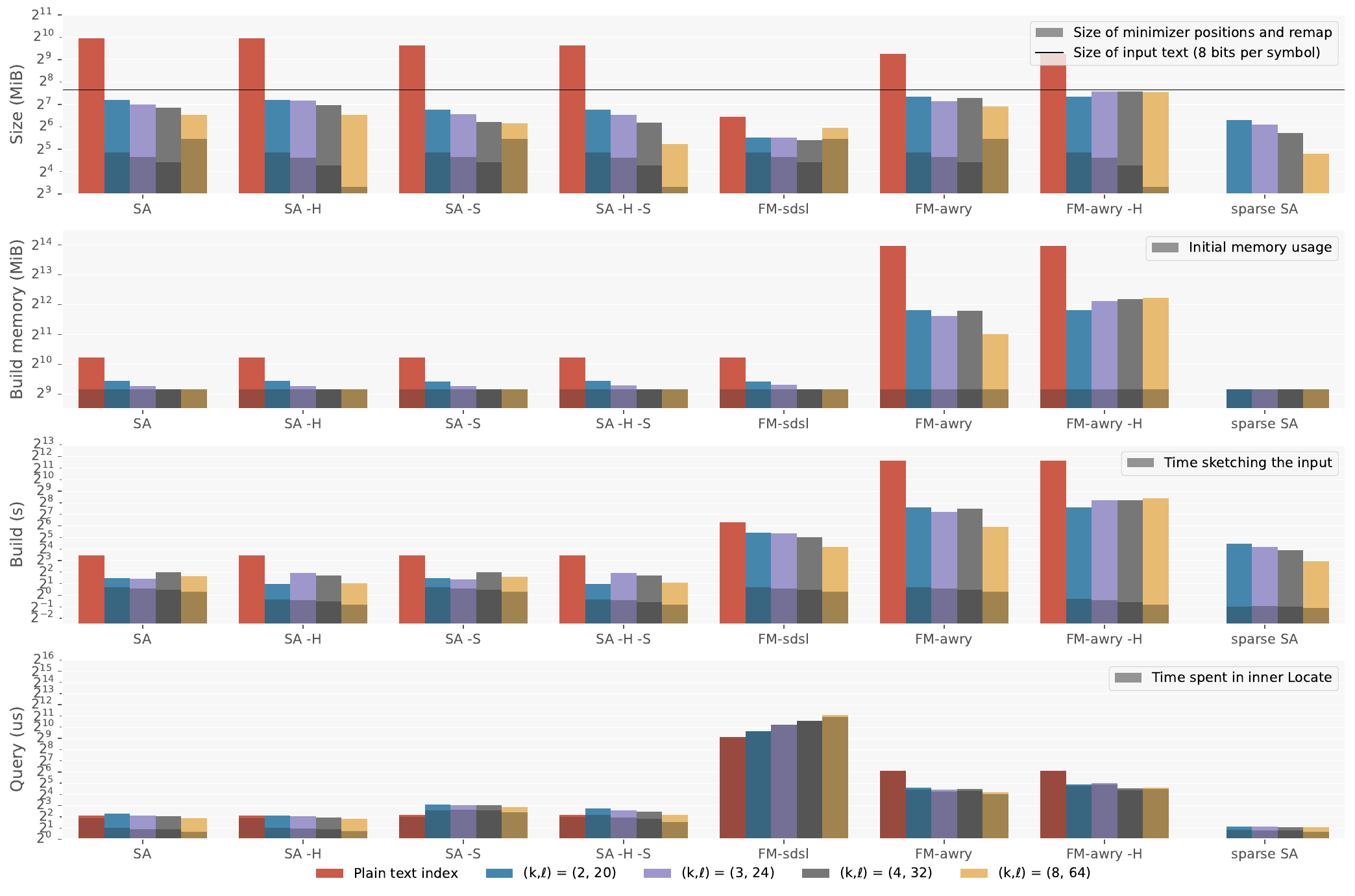}%
\caption{Results on $200$ MiB of English text.
Refer to the caption of \cref{fig:plot-v2}.}
\label{fig:plot-english}
\vspace{-14em} 
\end{figure*}


\end{document}